\documentclass[aps,prl,twocolumn]{revtex4}
\usepackage{amsmath}
\usepackage{txfonts}
\usepackage[scaled]{helvet}
\usepackage[pdftex]{graphicx}
\usepackage{float}

\usepackage{perpage}

\usepackage{color}
\usepackage{rotating}
\usepackage[breaklinks=true,colorlinks,citecolor=blue,linkcolor=blue,urlcolor=blue]{hyperref}
\usepackage{appendix}
\usepackage{graphicx}
\usepackage{natbib}
\usepackage{dcolumn}
\usepackage{bm}
\usepackage{hyperref}
\usepackage[mathlines]{lineno}
\usepackage[caption=false]{subfig}
\usepackage{epstopdf}

\begin{document}

\title{Spin-Valley Polarized Quantum Anomalous Hall Effect and a Valley-Controlled
       Half Metal in Bilayer Graphene}
\author{Xuechao Zhai$^{1,2}$, Yaroslav M. Blanter$^2$}
\affiliation{$^1$New Energy Technology Engineering Laboratory of
Jiangsu Province $\&$ School of Science, Nanjing University of Posts
and Telecommunications (NJUPT), Nanjing 210023, China\\
$^2$Kavli Institute of NanoScience, Delft University of Technology,
2628 CJ Delft, The Netherlands}

\date{\today}

\begin{abstract}
We investigate topological phases of bilayer graphene subject to
antiferromagnetic exchange field, interlayer bias, and irradiated by
light. We discover that at finite bias and light intensity the
system transitions into a previously unknown spin-valley polarized
quantum anomalous Hall (SVP-QAH) insulator state, for which the
subsystem of one spin is a valley Hall topological insulator (TI)
and that of the other spin is a QAH insulator. We assess the TI
phases occurring in the system by analytically calculating the
spin-valley dependent Chern number, and characterize them by
considering edge states in a nanoribbon. We demonstrate that the
SVP-QAH edge states lead to a unique spin rectification effect in a
domain wall. Along the phase boundary, we observe a bulk half-metal
state with Berry's phase of $2\pi$.
\end{abstract}

 \maketitle

\section{I.~~Introduction}

Bernal-stacked bilayer graphene (BLG) consists of two honeycomb
lattices coupled by van der Waals interaction \cite{Ohta}. Due to
its novel electronic properties, such as low-energy states with
Berry's phase of \cite{NovMc} $2\pi$ and an electrically-tunable
band gap \cite{McCann,ZhaTan}, BLG has recently attracted much
attention both theoretically and experimentally, see {\em e.g.}
Refs. \onlinecite{MarBlan,WeAll,JrJing,McKosh,RenQiao,JuWang}. In
particular, BLG became popular for studies of phenomena related to
the valley degree of freedom
\cite{RenQiao,ZhangJung,ZhangMac,LiZhang,JaskPelc}, resulting in
discovery of non-trivial topological states of matter. Recently, a
striking finding was the discovery of the quantum valley Hall (QVH)
effect in BLG, for which the edge states in the gap driven by an
interlayer bias are valley-polarized and provide the conductance
quantization with the step of \cite{JuShi,LiWang,YinJiang,JiangShi}
$4e^2/h$. Theoretical studies have predicted that the usual QVH
insulator in BLG undergoes a transition to a valley-polarized
quantum spin Hall insulator under effect of Rashba spin-orbit
interaction \cite{QiaoTse,ZhaiJ}.

Another attractive topological state of matter is the quantum
anomalous Hall (QAH) state, generally characterized by chiral edge
states in the gap and nonzero charge Chern numbers
\cite{ChangZh,Ezawa1,ZhaiJin}. The QAH state in BLG has been
suggested to be induced by Rashba interaction and internal
magnetization \cite{TseQiao} or by strong circularly polarized light
\cite{MohanRao,QuZhang,LagoMo}, differently from the mechanism of
Landau-level quantization in quantum Hall effect. Most recently, the
valley-polarized QAH state possessing the properties of both the QVH
and QAH states has been proposed in other 2D systems
\cite{PanLi,ZhSun} similar to BLG. Interestingly, this QAH state can
possess edge states which are not complectly chiral \cite{PanLi}.

In this work, we present the light and voltage controlled phase
diagram of BLG in contact with layered antiferromagnetic (LAF)
exchange field and discover a variety of topological insulator (TI)
states in this system. We argue that this AFM BLG itself is a
quantum anomalous valley Hall (QAVH) insulator \cite{CarSor}, that
it undergoes a transition to a usual QVH insulator as interlayer
bias increases, and that it is switched to a normal QAH (N-QAH)
insulator, a system with chiral edge states, as light intensity
increases. Remarkably, we discover that if both voltage and light
are applied, the system can be in a spin-valley polarized QAH
(SVP-QAH) insulator state, for which the subsystem of one spin is in
a QVH state and that of the other spin is in a QAH state. To our
knowledge, this state has not been previously described in the
literature and it is distinct from the single-valley spin-polarized
QAH states \cite{QuZhang,ZhSun} and the QAH states with no spin or
valley polarizations
\cite{PanLi,ZhangJung,LiZhang,ChangZh,Ezawa1,ZhaiJin,
JuShi,LiWang,YinJiang}.

The TI states in the system are discriminated by the spin-valley
dependent Chern number \cite{RenQiao}, which we calculate
analytically. We further characterize these TIs by edge states in a
zigzag ribbon and show transport differences in domain walls (DWs).
In the phase boundary, we discover that the 2D bulk system can host
a half-metal state with Berry's phase of $2\pi$, and the number of
gap-closing valleys is tunable.

Our paper is organized as follows. In Sec. II, we introduce the
system Hamiltonian and topological theory. In Sec. III, we show the
phase diagram and give the low-energy formulism. In Sec. IV, we
analytically derive the Chern numbers and show the domain-wall (DW)
transport. In the final section, we present the discussion and
conclusions.

\section{II.~~System Hamiltonian and Topological Theory}

\begin{figure}
\centerline{\includegraphics[width=6.0cm]{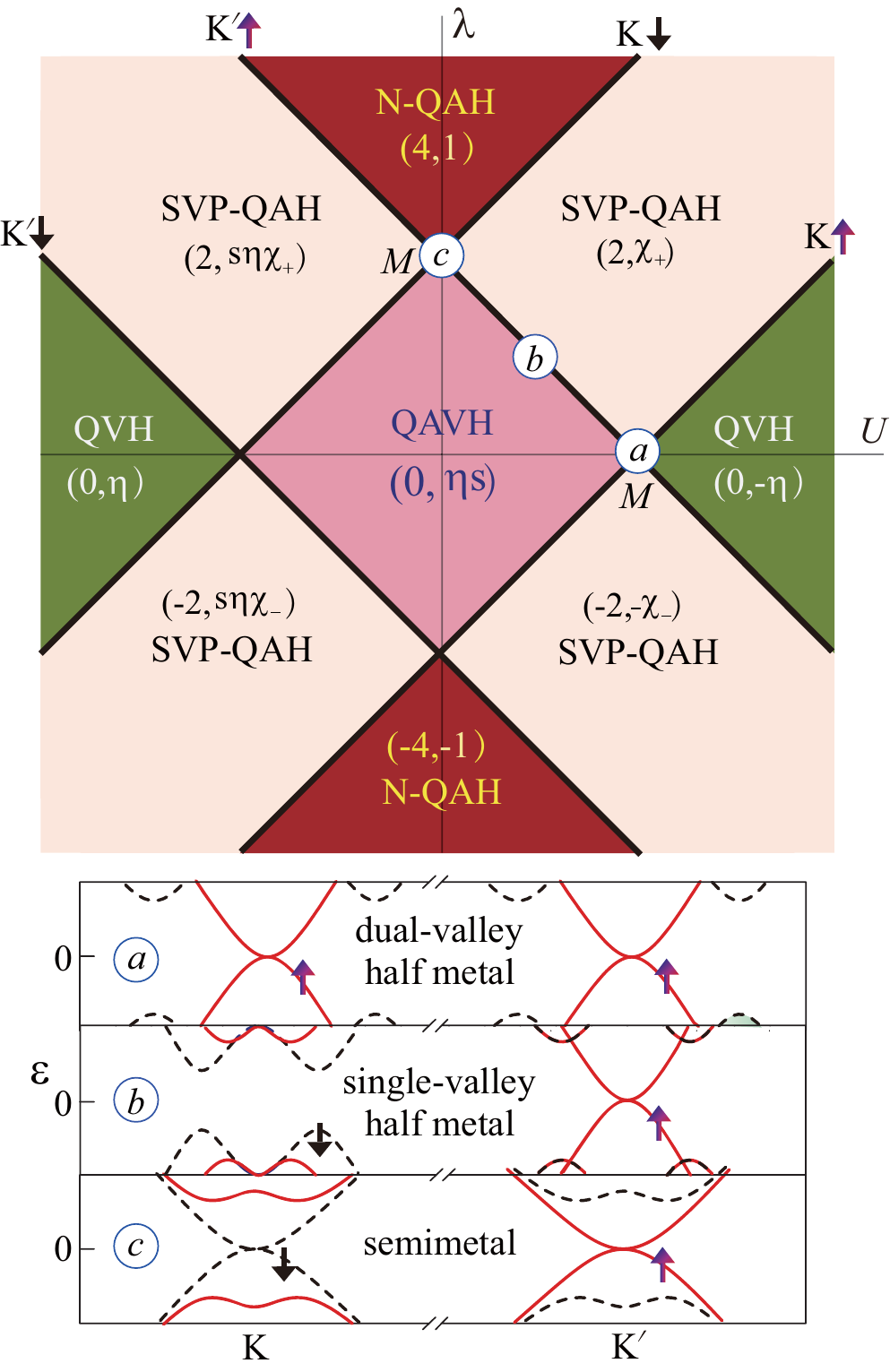}} \caption{(Color
online) Phase diagram in the $U$-$\lambda$ plane (top). The Chern
numbers $({\cal C},{\cal C}_\eta^s)$ are used to index the TI
phases. For the SVP-QAH state, ${\cal
C}_\eta^s=\chi_\pm=(\delta_{s,1}+\delta_{s,-1})(\delta_{s,1} \pm
s\eta\delta_{s,-1})$ [see Eq.~(11)]. Bold lines represent the phase
boundaries, where gap closing is indexed by $K$~($K'$) and
$\uparrow(\downarrow)$. The bulk bands for points $a$-$c$ are shown
at the bottom, where solid (dashed) lines denote
$\uparrow(\downarrow)$ subbands.}
\end{figure}

The BLG system we consider is described by the eight-band
tight-binding model
\begin{equation} \label{Hamiltonian}
\begin{split}
\hat{H}=&-J\sum_{{{\langle i,j\rangle}_{\|}}\alpha}c_{i\alpha}^\dag
c_{j\alpha}-\gamma\sum_{{{\langle
i,j\rangle}_{\bot}}\alpha}c_{i\alpha}^\dag c_{j\alpha}+
M\sum_{i\alpha}\mu_ic_{i\alpha}^\dag \sigma_zc_{i\alpha}\\
&-U\sum_{i\alpha}\mu_ic_{i\alpha}^\dag
c_{i\alpha}+i\frac{\lambda}{3\sqrt3}\sum_{{\langle\langle
i,j\rangle\rangle}_{\|}\alpha}\nu_{ij}c_{i\alpha}^\dag c_{j\alpha},
\end{split}
\end{equation}
where $c_{i\alpha}^\dag$ creates an electron with spin polarization
$\alpha$ at site $i$, and $\langle
i,j\rangle$/${\langle\langle}i,j{\rangle\rangle}$ run over all the
nearest and next nearest neighbor hopping sites. The subscript
${\|}~(\perp)$ denotes ``in plane'' (``out of plane''), $\mu_i=\pm1$
indicates that site $i$ lies in the bottom (top) layer, $\sigma_z$
is the $z$-component of Pauli matrix in the spin subspace, and
$\nu_{ij}=+1~(-1)$ holds if the next nearest hopping is
anticlockwise (clockwise). In Eq.~(\ref{Hamiltonian}), the first and
second terms are the intralayer and interlayer nearest neighbor
hoppings with \cite{McKosh} $J=3.0$~eV and $\gamma=0.39$~eV. The
third term, $H_M$, is the LAF exchange field, which can be applied
by depositing BLG between 2D magnetic insulators CrI$_3$
\cite{CarSor} ($M\sim$7~meV can be achieved). The fourth term,
$H_U$, is the interlayer bias with $U$ being up to 125~meV in
experiment \cite{ZhaTan,JrJing}. The fifth term, $H_\lambda$, is the
Haldane mass term \cite{Haldane}, which can be applied
\cite{QuZhang} by irradiating BLG with circularly polarized light
[see Supplemental Material (SM) \cite{SupMater} and
Refs.~\cite{KitaBra,Ezawa2}]. Estimated from the band width in BLG,
the light frequency would typically lie in the range
\cite{QuZhang,SupMater} $10^{13}$--$10^{14}$ Hz, leading to the
estimate of $\lambda\sim$10~meV.

The QAH effect is characterized by the charge Chern number
\cite{PanLi,TseQiao,ChangZh,Ezawa1,ZhaiJin}
\begin{equation}
{\cal C}=\frac{1}{2\pi}\sum_n\int_{\rm BZ}d^2{\bm k}\Omega_n(\bm k),
\end{equation}
where $\Omega_n$ is the momentum-space Berry curvature
\cite{RenQiao} in the out-of-plane direction for the $n$-th subband
\begin{equation}
\Omega_n(\bm k)=-\sum_{n'\neq n}\frac{2{\rm
Im}\langle\psi_{nk}|\upsilon_x|\psi_{n'k}\rangle
\langle\psi_{n'k}|\upsilon_y|\psi_{nk}\rangle}
{\left(\varepsilon_{n'}-\varepsilon_n\right)^2}.
\end{equation}
The summation is over all occupied valence bands in the first
Brilloin zone below the bulk band gap, $\psi$ is the Bloch state,
and $\upsilon_{x(y)}$ is the velocity operator along the $x(y)$
direction.

One can label a state a TI if it has at least one nonzero
topological invariant
\cite{TseQiao,QuZhang,ZhSun,CarSor,Haldane,KitaBra,Ezawa2,MohanRao,LagoMo}.
Because electrons in BLG here have a mixture of charge, spin and
valley degrees of freedom, the index $\cal C$ can not identify the
spin or valley polarized TI phases. To solve this problem, we employ
a decomposed, spin and valley dependent Chern number
\cite{RenQiao,Ezawa2} ${\cal C}_\eta^s$, which is well defined in
half of the first Brilloin zone around $K$ ($\eta=+1$) or $K'$
($\eta=-1$) points for spin-up ($s=+1$) or spin-down ($s=-1$)
electrons. On basis of ${\cal C}_\eta^s$, the spin and valley Chern
numbers \cite{RenQiao} can be defined as ${\cal C}_S={\cal
C}_\uparrow-{\cal C}_\downarrow=\sum_\eta({\cal
C}_\eta^\uparrow-{\cal C}_\eta^\downarrow)$ and ${\cal
C}_\upsilon={\cal C}_K-{\cal C}_{K'}=\sum_s({\cal C}_K^s-{\cal
C}_{K'}^s)$, respectively. The charge Chern number satisfies ${\cal
C}=\sum_{\eta,s}{\cal C}_\eta^s$.

\section{III.~~Phase diagram and Low-energy Formulism}

We present our result as the phase diagram in Fig.~1 by tuning $U$
and $\lambda$. The topological numbers $({\cal C},{\cal C}_\eta^s)$
are used to index the TI phases, including QAVH, QVH, N-QAH and
SVP-QAH. We show how to derive the analytical expressions for the
phase boundary and the Chern numbers from low-energy theory later.
Because the sign changes of $U$ and $\lambda$ only influence the
signs of Chern numbers (no novel phases appear), one only needs to
characterize the phases for ($U\geq0$, $\lambda\geq0$). To describe
the edge states, some typical cases are chosen to calculate the
bands in a zigzag ribbon (SM \cite{SupMater}). The edge states in
real space for each TI are illustrated in Figs.~2(a)-2(d), where the
moving directions (hollow arrows) for electrons agree with
Eqs.~(\ref{Chern_index_TI}-\ref{Chern_total}). At the phase
boundary, the bulk gap closes [Eq.~(\ref{phase-boundary})].

We first discuss the QAVH state, for which $({\cal C},{\cal
C}_\eta^s)=(0,\eta s)$ \cite{CarSor}. The nonzero value of ${\cal
C}_\eta^s$ in this phase arises from the inversion symmetry breaking
dominated by $M$. Although the QAVH phase has metallic in-gap edge
states near $K$ and $K'$, the index $\cal C$ is zero. The anomalous
property of the state is that the subsystem of each spin is a QVH
insulator [${\cal C}_\upsilon^\uparrow={\cal C}_K^\uparrow-{\cal
C}_{K'}^\uparrow={\cal C}_{K'}^\downarrow-{\cal
C}_K^\downarrow=-{\cal C}_\upsilon^\downarrow=2$ in Fig.~2(a)],
while the whole system is not valley polarized (${\cal
C}_\upsilon=0$). The result ${\cal C}_K^\uparrow+{\cal
C}_K^\downarrow={\cal C}_{K'}^\uparrow+{\cal C}_{K'}^\downarrow=0$
indicates the edge states in each valley behaves as a quantum spin
Hall state \cite{QiaoTse,Ezawa1} in spite of ${\cal C}_S=0$. If we
include non-zero voltage $U$, the system eventually undergoes a
transition to a normal QVH insulator [${\cal C}_\upsilon=-4$ in
Fig.~2(b)] due to the competition of $U$ and $M$. The QAVH insulator
transitions to an N-QAH insulator [${\cal C}=4$ in Fig.~2(c)] as
$\lambda$ increases due to the time-reversal symmetry breaking. Each
transition is associated with with a process of the closing and
reopening of the gap.

\begin{figure}
\centerline{\includegraphics[width=7.0cm]{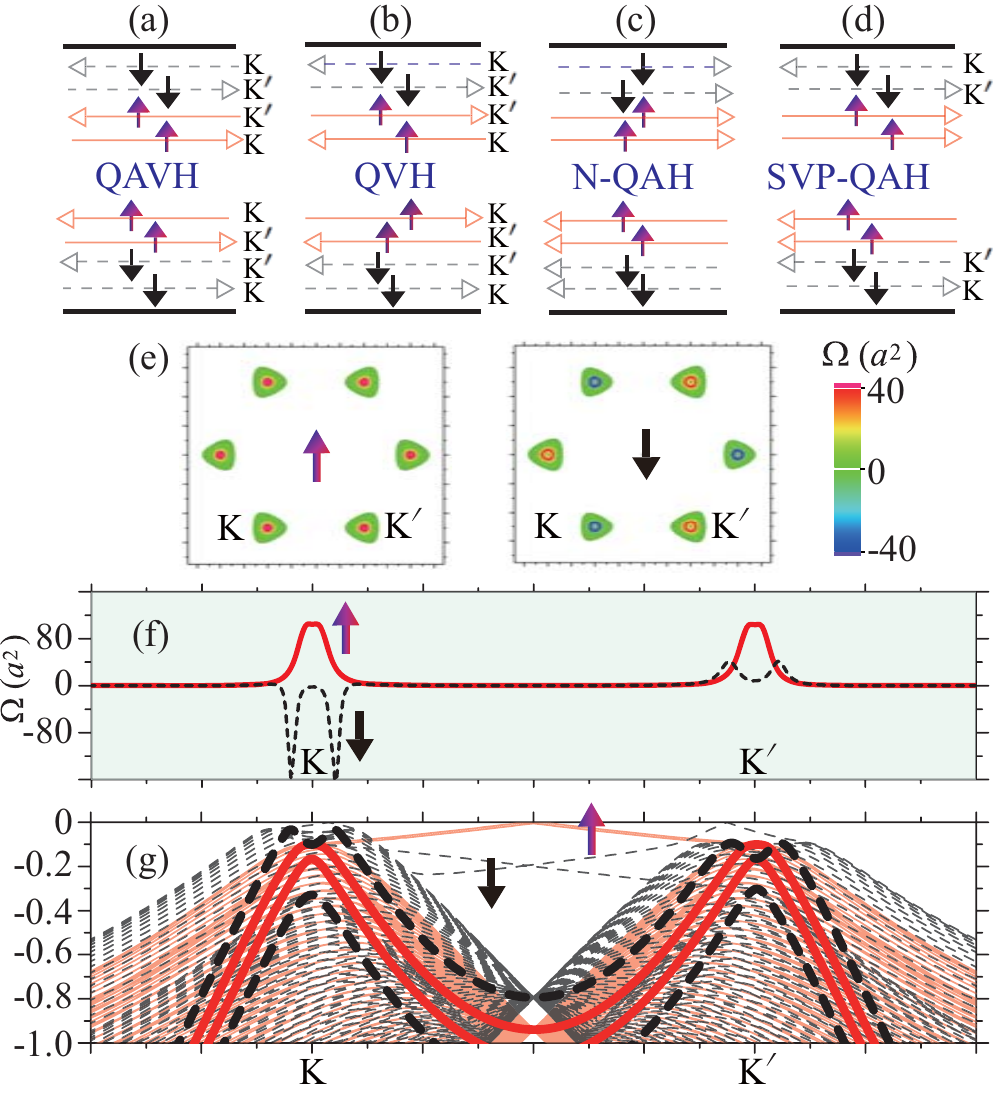}} \caption{(Color
online) (a)-(d) Schematics of eight edge states in the bulk gap for
four TIs in Fig.~1, corresponding to our calculated band structures
\cite{SupMater} of a zigzag-edged ribbon for typical points under
$(U\geq0,\lambda\geq0)$. (e)-(g) Electronic properties of the
SVP-QAH state. Berry curvature in (e) $(k_x,k_y)$ plane and (f)
along $K$-$K'$ direction, corresponding to (g) the bulk (thick
lines) and ribbon (thin lines) bands.}
\end{figure}

Our most significant result is the SVP-QAH phase, which appears in
regions where $U\lambda\neq0$ and occupies a major part of the phase
diagram. For a SVP-QAH state , four spin-down edge states in the gap
appear near $K$ and $K'$, but four spin-up edge states appear away
from $K$ and $K'$ [Figs.~2(d) and 2(g)]. Correspondingly, the Berry
curvature in the $(k_x,k_y)$ plane [Fig.~2(e)] and specifically
along $K$-$K'$ direction [Fig.~2(f)] is shown. We find that the
spin-down subsystem is a QVH insulator [${\cal
C}_K^\downarrow=-{\cal C}_{K'}^\downarrow=-1$, ${\cal
C}_\upsilon^\downarrow=-2$, ${\cal C}^\downarrow={\cal
C}_K^\downarrow+{\cal C}_{K'}^\downarrow=0$], whereas the spin-up
subsystem is a normal QAH insulator [${\cal C}_K^\uparrow={\cal
C}_{K'}^\uparrow=1$, ${\cal C}_\upsilon^\uparrow=0$, ${\cal
C}^\uparrow={\cal C}_K^\uparrow+{\cal C}_{K'}^\uparrow=2$], and then
${\cal C}=-{\cal C}_\upsilon={\cal C}_S=2$ is achieved. To our
knowledge, this SVP-QAH insulator is a new type of a QAH state,
which differs obviously from the single-valley spin-polarized QAH
\cite{QuZhang,ZhSun}, spin-unpolarized QAH
\cite{PanLi,RenQiao,ZhangJung,ZhangMac,LiZhang}, and
valley-unpolarized QAH \cite{Ezawa1,Ezawa2,RenQiao} states
discovered previously.

At the phase boundaries, see bold lines in Fig.~1, there exists a 2D
bulk state that is half-metallic \cite{SonCohen} (gap closing at $K$
or $K'$), as shown by the completely spin-polarized band structures
(bottom) near the Fermi level for the marked points $a$ and $b$. In
contrast, point $a$ corresponds to a dual-valley half-metal (valley
degeneracy), while point $b$ refers to a single-valley half-metal
(the other valley is insulating). For the marked point $c$,
spin-down (spin-up) subbands near the Fermi level are locked to the
$K$ $(K')$ valley, resulting in a spin-valley locked semimetal. Thus
gap closing of the valleys in which one or two is tunable. Notably,
the bulk half-metal state here has the Berry's phase of $2\pi$ [Eq.
(\ref{berry})].

To further explore the physics underlying the phase diagram, we
analyze the low-energy effective Hamiltonian derived from the
lattice model (1) after Fourier transformation, and reveal the
physics of electrons near $K$ and $K'$. For simplicity, we set
$\hbar=1$ and $\upsilon_{\rm F}=1$ below. Because two opposite spins
are decoupled in Eq.~(1), the effective Hamiltonian can be written
as a $4\times4$ matrix
\begin{equation} \label{Low-energy-Hamiltonian}
\begin{split}
{\cal H}_{\bm k}=\left(\begin{array}{cccc}
\Delta_{\eta s}^+-U, &\eta k_x- ik_y, &0, &0 \\
\eta k_x+ik_y, &-\Delta_{\eta s}^--U, &-\gamma, &0 \\
0, &-\gamma, &\Delta_{\eta s}^-+U, &\eta k_x- ik_y \\
0, &0, &\eta k_x+ik_y, &-\Delta_{\eta s}^++U \\
\end{array}\right),
\end{split}
\end{equation}
where $\Delta_{\eta s}^{\pm}=\eta\lambda\pm sM$. Note that $s$ and
$M$ only occur as a product $sM$, and thus the reversal of the sign
of $M$ is equivalent to the reversal of spin orientation. We will
assume $M > 0$. By matrix diagonalization, we derive the band
structure as
\begin{equation}
\begin{split}
\varepsilon&=\pm\sqrt{k^2+(sM-U)^2+\lambda^2+\frac{1}{2}(\gamma^2\pm\sqrt{\Gamma_k})},\\
\Gamma_k&=k^2[16(sM-U)^2+4\gamma^2]+[\gamma^2-4\eta\lambda(sM-U)]^2.
\end{split}
\end{equation}
Note that the electron-hole symmetry is present in Eq.~(5), from
which one can prove that the gap is closed only under
$\varepsilon=0$, which requires $k=0$ [at $K$($K'$)] and
simultaneously
\begin{equation} \label{phase-boundary}
U=sM+\eta\lambda.
\end{equation}
According to the topological theory \cite{RenQiao}, this gap-closing
condition in Eq. (\ref{phase-boundary}) precisely describes the
phase boundary as given in Fig.~1. Specifically at $k=0$, the energy
gap reads
\begin{equation}
\Delta_\eta^s=2\sqrt{(sM-U)^2+\lambda^2+\frac{1}{2}\left(\gamma^2-|\gamma^2
-4\eta\lambda(sM-U)|\right)}.
\end{equation}

Judging from Eq.~(\ref{phase-boundary}), the system is a dual-valley
half-metal for $(\lambda,U)=(0,\pm M)$ and a spin-valley locked
semimetal for $(\lambda,U)=(\pm M,0)$. Except for these points, the
system is a single-valley half-metal in the phase boundary. The
effective $2\times2$ Hamiltonian used in Ref.~\onlinecite{NovMc} can
be derived to describe the low-energy half-metal subbands with
parabolic dispersion. The half-metal Bloch state $\psi_{k,\phi}$
($\phi$ being the azimuthal angle of the momentum) here has the
Berry phase \cite{ZJ2014}
\begin{equation} \label{berry}
\varphi_{\rm
B}=i\int_0^{2\pi}d\phi\langle\psi_{k,\phi}|\frac{\partial}
{\partial\phi}|\psi_{k,\phi}\rangle=2\eta\pi,
\end{equation}
which can induce a spin-polarized integer quantum Hall effect with a
missing zero-level plateau \cite{NovMc}.

\section{IV.~~Chern numbers and domain-wall (DW) transport}

We see from Eq.~(\ref{phase-boundary}), Fig.~1 and our band
calculations \cite{SupMater} that the QAVH, QVH, N-QAH and SVP-QAH
insulator states hold under $|\lambda|<M-|U|$, $|\lambda|<|U|-M$,
$|\lambda|>|U|+M$, $\big||U|-M\big|<|\lambda|<|U|+M$, respectively.
Usually, the topological invariants are associated with the signs of
the parameters for a TI state \cite{RenQiao,Ezawa2}. For the QAVH,
QVH, and N-QAH insulators here, the index ${\cal C}_{\eta}^s$ is
respectively determined by
\begin{equation} \label{Chern_index_TI}
{\cal C}_{\eta}^s=\eta s,~-\eta{\rm sgn}(U),~{\rm sgn}(\lambda)
\end{equation}
and correspondingly ${\cal C}$, ${\cal C}_S$, ${\cal C}_\upsilon$
are expressed as
\begin{equation} \label{Chern_total_TI}
{\cal C}=0, 0,~4{\rm sgn}(\lambda)\parallel{\cal
C}_S=0,0,0\parallel{\cal C}_\upsilon=0,-4{\rm sgn}(U),0,
\end{equation}
where ${\rm sgn}(x)$ is the sign function. For the SVP-QAH
insulator, we calculate ${\cal C}_{\eta}^s$ to be
\begin{equation} \label{Chern_index_SVPQAH}
{\cal C}_{\eta}^s=\left(\delta_{s,1}+\delta_{s,-1}\right)\left[{\rm
sgn}(\lambda)\delta_{s,{\rm sgn}(U)}+s\eta\delta_{s,{\rm
sgn}(-U)}\right],
\end{equation}
and the other Chern numbers satisfy
\begin{equation} \label{Chern_total}
{\cal C}=2{\rm sgn}(\lambda),~{\cal C}_s=2{\rm sgn}(\lambda
U),~{\cal C}_\upsilon=-2{\rm sgn}(U).
\end{equation}
Here, $\delta_{\mu,\nu}$ is the Kronecker delta function. We see
from Eqs.~(\ref{Chern_index_TI}) and (\ref{Chern_index_SVPQAH}),
$|{\cal C}_{\eta}^s|=1$ always holds for each TI case.

\begin{figure}
\centerline{\includegraphics[width=8cm]{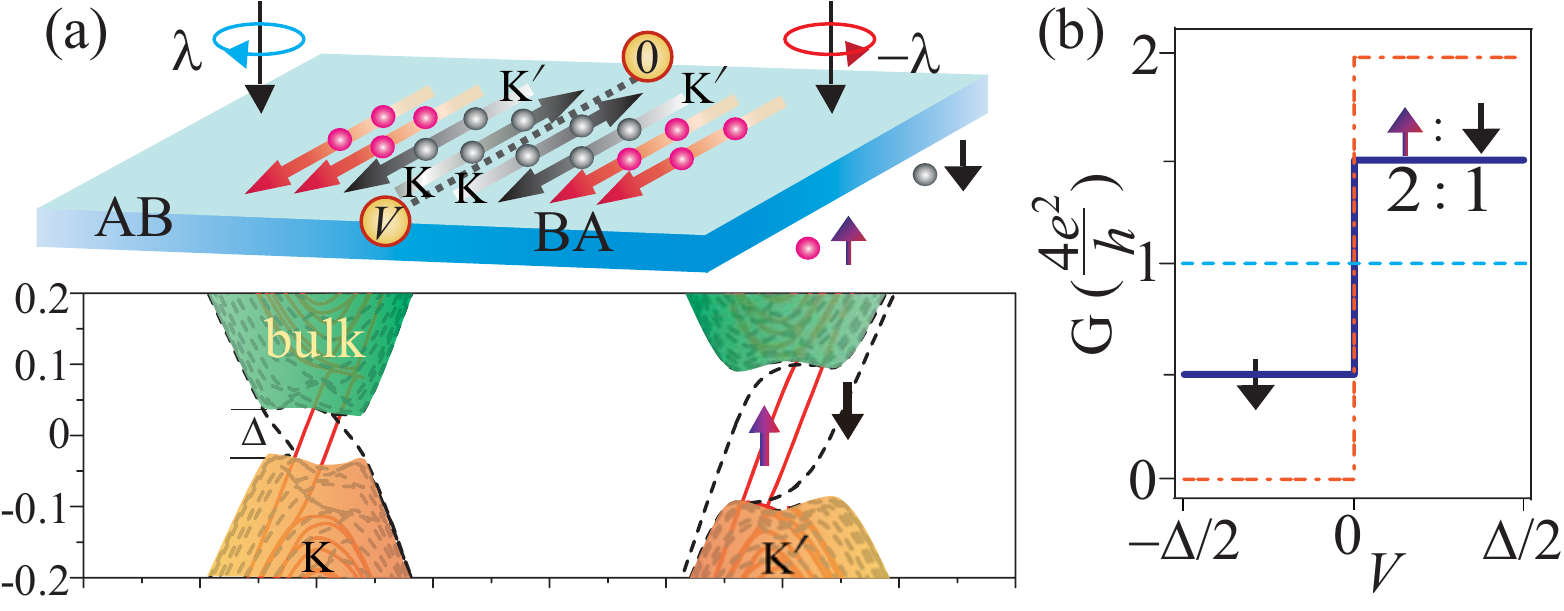}} \caption{(Color
online) (a) Sketch of a DW (top) between AB- and BA-stacked BLG. The
voltage $V$ (0) is for detection. The conducting channels are shown
for the AB$(\lambda)$/BA$(-\lambda)$ case, with the band (bottom)
given in a sharp DW. The band gap satisfies $\Delta\simeq{\rm
min}~\Delta_{\eta}^{s}(\eta,s=\pm1)$. (b) DW conductance within the
gap in the AB$(\lambda)$/BA$(-\lambda)$ DWs. The solid, dashed,
dot-dashed lines represent, respectively, the SVP-QAH, QAVH (the
same for QVH) and N-QAH cases.}
\end{figure}

To distinguish and detect the SVP-QAH state, we consider a DW
between AB- and BA-stacked GBL \cite{JuShi} in Fig.~3(a), where the
conducting channels and band structure are shown when $\lambda$ has
the opposite values in AB and BA regions. By using Green's function
method \cite{Datta}, we show the DW conductance [$G=I/V$, where
$I(V)$ is the current (bias)] in Fig.~3(b) and predict a unique spin
rectification effect (solid line) for the SVP-QAH edge states. We
assume here that the DW is much shorter than the electron mean free
path, and thus scattering is ignored. The ratio of channel numbers
for $\uparrow$ and $\downarrow$ is $2:1$ for $V>0$, while only
$\downarrow$ channels contribute to transport for $V<0$. In
contrast, no charge or spin rectifications occur for the QAVH or QVH
states. The N-QAH state acts as a charge diode, however, still no
spin rectification occurs.

The spin/valley filter or diode can be further realized by tuning
the types of DW and fields (SM \cite{SupMater}). Specifically, a
unipolar spin diode \cite{ZhaiWen} (only spin-up electrons are
unidirectionally conducting) works if the DW in Fig.~3(a) is of the
armchair type due to the absence of the QVH edge states
\cite{RenQiao,JuWang,JaskPelc,ZhangMac}.

\section{V.~~Discussion and conclusions}

Realization of the SVP-QAH state in BLG depends strongly on the LAF
field, $H_M$, which is not induced by the electron-electron
interaction \cite{QuZhang} but by proximity, such as a van der Waals
system consisting of BLG and magnetic insulator CrI$_3$
\cite{CarSor} ($M\sim7$ meV can be achieved). Possibilities of
proximity induced by other 2D magnets, such as CrBr$_3$ or
Cr$_2$Ge$_2$Te$_6$ \cite{GongZhang}, requires more research.
Magnetic insulator EuS is an alternative candidate to induce $H_M$
on basis of the relevant experimental evidence \cite{WeiLee}. Beyond
model~(\ref{Hamiltonian}), the Coulomb interaction might induce a
gap about $1$~meV \cite{JrJing,QuZhang}, which is usually
disregarded when other interactions are stronger
\cite{ZhaTan,JuShi,LiWang,YinJiang} as we assume. Notably, the LAF
order can not be easily broken by light in spite of magneto-optical
Kerr or spin-phonon coupled effects \cite{SunZ,HuangCenker}. If the
LAF system becomes ferromagnetic, {\it e.g.} by applying magnetic
field \cite{CarSor}, the SVP-QAH phase disappears.

In particular, the time-periodic optical field can be safely reduced
to the time-averaged static term $H_\lambda$ under high-frequency
($>$3J) condition \cite{Ezawa2,LagoMo,MohanRao,QuZhang}, when the
Chern number is well defined to characterize the quantized current
\cite{DehghOka,Kundu,Manghi} of edge states near $E=0$. This is
further carefully checked (SM \cite{SupMater}) by using the
full-time Floquet theory. Because the light absorbance per graphene
layer is only about 2.3\% \cite{MakSfe}, the band population is
presumably intact. For the SVP-QAH DW transport in Fig.~3(a), the
electron mean free path $L_0\sim420$~nm was found by experiment
\cite{JuShi,YinJiang,LiWang,JiangShi}. Ignoring the contact
resistance, the dependence of ballistic conductance $G$ on DW length
$L$ reads \cite{Datta,JuShi} $G=(\zeta e^2/h)(1+L/L_0)$, where
$\zeta$ is the integer for quantized plateau in Fig.~3(b). For the
lower-frequency light, other complex effects such as
\cite{FoaPerez,SenCla,WangStein} time-averaged densities of states
in addition to edge states can break the conductance quantization
\cite{SupMater}.

{\color{red}The exchange proximity effect in Graphene/2D-magnet may
be rather small that pressure is usually needed to induce a
measurable QAH effect \cite{ZhangZhao}. Certainly, small details
related to the physics close to the Fermi level such as bands
hybridization, charge transfer, exchange proximity or bias effects
may change the phase diagram at the level of ab-initio calculations
\cite{CarSor}, reduce the conductivity and even hamper the
observation of the SVP-QAH phase. Therefore, more efforts both in
theory and experiment are still needed to explore and confirm the
SVP-QAH phase in GBL/2D-magnet systems.}

Another striking result revealed by Eq.~(11) is that, without
optical fields, the SVP-QAH phase holds in the systems with
intrinsic spin-orbit coupling (by replacing $H_\lambda$ with
$sH_\lambda$). A promising alternative system to obverse the SVP-QAH
phase is the experimentally-available LAF Eu/bilayer silicene/Eu
structure \cite{TokAver}. Our further calculations \cite{SupMater}
indicate the SVP-QAH is robust against weak Rashba interactions.

In summary, the concept of van der Waals LAF order applied to 2D
valleytronics, rarely noticed before, opens a novel avenue for
engineering small devices with ultra-low dissipation. The explored
topological transitions modulated by photon and electric voltage
will motivate the experimental exploration of novel 2D QAH
insulators with rich spin or valley physics in the integrated
systems between layered 2D materials, van der Waals magnets and
correlated valley materials, for which the combination of
spintronics, valleytronics and topology promises discoveries of
novel condensed matter phases.

\section{Acknowledgments}

This work was supported by the NSFC with Grant No. 61874057, QingLan
Project of Jiangsu Province (2019), 1311 Talent Program ``DingXin
Scholar" of NJUPT (2018), and Jiangsu Government Scholarship for
Overseas Studies.

\end{document}